\documentclass[aps,pra,twocolumn,floatfix,a4paper,nopacs,superscriptaddress]{revtex4}
\usepackage{epsfig}
\usepackage{amsmath}
\usepackage{amssymb}
\usepackage{color}

\newcommand{\ii}{\mathrm{i}}
\newcommand{\e}{\mathrm{e}}

\newcommand{\ke}[1]{\left|{#1}\right>}
\newcommand{\ket}[1]{\ke{#1}}
\newcommand{\bra}[1]{\left<{#1}\right|}

\newcommand{\id}{\mathbb I}

\newcommand{\subspace}{{\mathcal S}}
\newcommand{\tr}{\mathrm{Tr}\,}

\newcommand{\hi}{\mathcal{H}}
\newcommand{\dep}{\mathcal{D}}
\newcommand{\uni}{\mathcal{U}}

\newcommand{\spann}{\mathrm{span}\,}
\newcommand{\va}{\mathbf{a}}
\newcommand{\vb}{\mathbf{b}}
\newcommand{\diag}{\mathrm{diag}}

\begin{document}
\title{Grover search under localized dephasing}

\author{Daniel Reitzner}
\affiliation{Research Center for Quantum Information, Institute of Physics, Slovak Academy of Sciences, D\'ubravsk\'a cesta 9, Bratislava 84511, Slovakia}
\affiliation{Faculty of Informatics, Masaryk University, Botanick\'a 68a, 60200 Brno, Czech Republic}
\author{Mark Hillery}
\affiliation{Department of Physics, Hunter College of the City University of New York, 695 Park Avenue, New York, New York 10065 USA}

\begin{abstract}
Decoherence in quantum searches, and in the Grover search, in particular, has already been extensively studied, leading very quickly to the loss of the quadratic speedup over the classical case, when searching for some target (marked) element within a set of size $N$. The noise models used were, however, almost always global. In this paper, we study Grover search under the influence of localized partially dephasing noise of rate $p$. We find that, in the case when the size $k$ of the affected subspace is much smaller than $N$, and the target is unaffected by the noise, namely when $kp\ll\sqrt{N}$, the quadratic speedup is retained. Once these restrictions are not met, the quadratic speedup is lost. If the target is affected by the noise, the noise rate needs to scale as $1/\sqrt{N}$ in order to keep the speedup. We also observe an intermediate region, where if $k\sim N^\mu$ and the target is unaffected, the speedup seems to obey $N^\mu$, which for $\mu>0.5$ is worse than the quantum, but better than the classical case. We also put obtained results for quantum searches into perspective of quantum walks and searches on graphs.
\end{abstract}

\maketitle

\section{Introduction}

Grover search \cite{Grover97}  is one of several quantum algorithms that provide us with speedups when compared with classical counterparts. Its basic function is searching over a database of $N$ elements in which no prior structure between the database elements is known. In this setting, quantum mechanics offers quadratic speedup over classical (blind) search; Grover search finds marked element in $O(\sqrt{N})$ steps, while in the classical case we need $O(N)$ steps.
It has been proven, that the quadratic speedup is optimal under assumption of unitary evolution \cite{BeBeBrVa97,BoBrHoTa98,Zalka99} with Grover algorithm reaching optimal scaling---we will call this specific evolution \emph{Grover evolution}.


Since the unitarity is only an idealized situation, quantum searches have also been studied under various models of decoherence to determine their functionality under these more realistic conditions. A recurring observation is that the quadratic speedup is quickly lost; the decoherence rate for which this happens is usually of order $1/\sqrt{N}$.
Studying evolutions under decoherence is typically difficult, because one has to overcome the difficulties arising from the departure from pure states. In particular, the usual approach in the unitary case---identification of invariant subspaces---is no longer applicable and different methods have to be employed.

The rapid loss of efficiency is observed not only for Grover search under various types of decoherence \cite{PaRu99,LoLiZhTu00,Azuma02,ShMoBi03,ShBrWh03,HsLiCh04,MaEtal06,CoDeBuDe16} but also in more general quantum search scenarios where the Grover evolution is not set and, for purpose of the search, any evolution with an arbitrarily large ancillary system that undergoes some decoherence is considered \cite{ReSc08,Temme14,VrReReWo14}. 
The works studying noisy quantum searches typically apply the decoherence on the whole Hilbert space where evolution happens. This includes noisy oracle \cite{ShBrWh03,ReSc08,Temme14} or global qubit register decoherences \cite{Azuma02,ShMoBi03,CoDeBuDe16} in various forms, such as
depolarization, dephasing, or some deviations from unitary Grover evolution.

To our knowledge, localized decoherence in a quantum search has been studied only in Ref.~\cite{rastegin}, which treats a specific case of what we do here.  Whereas our results are only approximations, those presented in the reference are exact, and we will later make a comparison between the two works. Results of Ref.~\cite{nahimovs} are also somewhat related---their findings show that in the case of multiple marked elements, some of which are faulty, the non-faulty ones can still be efficiently found. Our paper does not study this situation, as we consider only single marked vertex, but their results are in line with our general observation that if the marked vertex is faulty, it cannot be efficiently found.

In this paper, we are motivated by a situation in which only a subspace of the Hilbert space is affected---this might happen, e.g., if the database is encoded as a (multiple-)qubit system, where only one qubit (or a few) undergo decoherence. Such a model was outlined in Refs.~\cite{Azuma02,ShMoBi03,CoDeBuDe16}. There the authors studied a situation where the search is performed on a system encoded in qubits and these undergo decoherence individually. However, in all these works, the locally applied noise was added to all the qubits and so the localization of it was not studied.

In this paper, we present an approach to the problem of localized decoherence in the Grover evolution. We focus on the question of whether the limited localized influence of the decoherence can loosen the strict bounds on the speedup. We will make use of a method which is based on identifying invariant subspaces in which evolution takes place---we will identify invariant subsace not in the underlying Hilbert space but in the linear vector space of specific operators. Whereas in the unitary evolution these subspaces had an operational meaning of state subspaces through which the evolution was defined, in our case the invariant subspace is only an abstract mathematical construct where the usual interpretations no longer hold. Nevertheless, the approach we use in the end provides tangible results.

The paper is organized as follows. In the next, section we provide a Grover algorithm for reference and setting the notation. In Sec.~\ref{sec:noise}, we define the noise model and provide technical details on how it affects density matrices. In Sec.~\ref{sec:invariant}, we define the method of invariant subspaces on density matrices fit to the problem of Grover evolution. Results are presented in the Sec.~\ref{sec:results} for several scenarios. As in recent years, quantum searches have also been heavily studied in quantum walks setups, in Sec.~\ref{sec:qws} we alter the methodology to suit the framework of quantum walks which we also introduce in that section. The conclusions of the paper are provided in Sec.~\ref{sec:conclusion}.

\subsection{Invariant subspace in the Grover search}
\label{sec:GS}

Formally, the Grover algorithm allows for searches on Hilbert space $\hi$ of dimension $N$ for an element marked by a (quantum) oracle, which is represented by a unitary operation
\begin{equation}
O_f: \ket{x}\otimes\ket{m} \mapsto \ket{x}\otimes\ket{m \oplus f(x)},
\end{equation}
defined on elements of \emph{the canonical basis} $\ket{x}\in\hi$, $\ket{m}\in\mathbb C^2$ and $\oplus$ is addition modulo 2. Boolean function $f$ is the classical oracle associated with $O_f$---in a sense, in the quantum oracle there is only as much information as in the classical oracle with the difference being that the quantum oracle can work also with non-classical state on its input. Since such oracle provides no further information about the structure of the marked element(s), it is called unstructured. In this paper, we will use only such oracle.

The Grover algorithm makes use of two operators. One is derived from the oracle $O_f$---by observing that states $\ket{x}\otimes\ket{-}$ with $\ket x$ from the canonical basis are eigenstates of $O_f$ with eigenvalues $\pm1$, we define new oracle $R_f$ as the action of oracle $O_f$ on mentioned state. This allows us to drop the second part of the state and write the action of the oracle simply without the ancillary system as
\begin{equation}
R_f:\ket{x} \mapsto (-1)^{f(x)}\ket{x}.
\end{equation}
The second operator used in the Grover search is \emph{the inversion about average,}
\begin{equation}
\label{eq:G}
G=2\ke{s}\bra{s}-\id,
\end{equation}
where
\begin{equation}
\label{eq:groverinit}
\ket{s}=\frac{1}{\sqrt N}\sum_{x}\ket x
\end{equation}
is the equal superposition over all canonical states. Note that if we define $t=2/N$ and $r=1-t$, for each canonical state $\ket x$ we have $G\ket x=-r\ket x+t\sum_{y\neq x}\ket y$.

As we will not deal with more than one marked element in this paper, we suppose now that the oracle marks only a single element. Then the result of Ref.~\cite{Grover97} is that by defining $U=GR_f$, one can express the success probability after $m$ steps of evolution by formula
\begin{equation}
\label{eq:GSsuccess}
p_{\mathrm{suc}}(m)=\sin^2\left[(2m+1)\frac{\theta}{2}\right],
\end{equation}
where $\cos\theta=r$.
This probability is maximized when $(2m_0+1)\theta=\pi$, which gives the optimal number of steps,
\begin{equation}
\label{eq:m0}
m_0\simeq\frac{\pi}{4}\sqrt{N},
\end{equation}
valid for a large number of elements $N$. This number of steps will appear multiple times in the rest of the paper, where it will always be denoted as $m_0$. 
This number of steps implies that one
needs only $m_0\sim O(\sqrt{N})$ repetitions of $U$, i.e., calls to the oracle $R_f$, in order to transform the initial state $\ket s$ into the  state
\begin{equation}
\ket{e}=\ket{f^{-1}(1)},
\end{equation}
which is the marked element, or \emph{the target.} This means that the quantum search is quadratically faster than the best classical search, which requires $N/2$ queries to the oracle $f$ on average. 

The states $\ket e$ and $\ket s$ define an invariant subspace in which the evolution happens. Specifically, defining $\subspace=\spann\{\ket{s},\ket{e}\}$, for any $\ket{\psi}\in\subspace$ also $U\ket{\psi}\in\subspace$. This method is also employed in more involved cases, commonly in quantum walks, where a precise identification of invariant subspaces is essential and leads to similar speedups; see, e.g., Refs.~\cite{krovi,ReHiFeBu09} for graph-specific definition (see also Sec.~\ref{sec:qws} of this paper).

As a side note, let us mention that for the purpose of estimating the efficiency of the search, the requirement on the oracle is just its computational complexity being low. For example, in the classical case of an unstructured search one can think of a search for a name belonging to a known number. The phone book works as an oracle---it takes $O(\log N)$ steps to find whether a queried person belongs to the particular number if the phone book has $N$ entries). Similarly, in the quantum case the oracle may be just a subroutine of a more complex algorithm, such as in the case of the algorithm for element distinctness in Ref.~\cite{Ambainis07}.

\section{Noise model}
\label{sec:noise}

Global noise in Grover search, and in quantum searches in general, has a strong degrading effect on the efficiency of the algorithm. Typically, already with noise rates stronger than $1/\sqrt{N}$ the quadratic speedup is lost and quantum searches offer only linear speedup at best. The question stands whether this undesirable property can be lifted if we consider only localized noise, as it seems unrealistic for a noise to be dependent (in this way) on the number of the database elements.

A consequence of the fact that the noise destroys coherence in the system is that we need to switch from the pure state formalism to the density matrix formalism, in which the state is described by a trace-one operator, typically labeled $\varrho$. The unitary evolution described in the previous section now reads $\uni(\varrho)=U\varrho U^\dagger$. The noise $\dep_p$ is parametrized by its strength $p\in[0;1]$ and will affect the state between any two applications of the unitary. The evolution of the state $\varrho$ under decoherence will now be described as
\begin{equation}
\label{eq:generalevolution}
\varrho(m)=\left(\uni \circ \dep_p\right)^m(\varrho),
\end{equation}
where $m$ is the number of performed steps of the evolution and the ``exponentiation'' is in the sense of concatenation of the operations. This type of evolution is standard in the literature. Here, however, the noise $\dep_p$ will not affect the whole Hilbert space $\hi$, but rather only a small subset of it.

For such a noise model, we consider partially dephasing localized noise, where the dephasing is defined via partial projection into a subspace of a Hilbert space. More concretely, let us consider a Hilbert space $\hi$ of dimension $N$ and suppose we can split it into two subspaces $\hi_0$ and $\hi_1$ such that $\hi=\hi_0\oplus\hi_1$ with the dephasing acting on the whole subspace $\hi_0$. Denoting the projection onto $\hi_0$ as $\Pi_0$ and the projection onto the orthogonal complement as $\Pi_0^\perp=\id-\Pi_0$, the partial dephasing with rate $p$ has the form
\begin{equation}
\label{eq:noise}
\dep_p(\varrho)=p\Pi_0\varrho\Pi_0+p\Pi^\perp_0\varrho\Pi^\perp_0+(1-p)\varrho.
\end{equation}
Note that since $\Pi_0^\perp=\Pi_1$ is a projection onto $\hi_1$, the previous equation treats the two subspaces symmetrically.
This noise thus performs dephasing between the two subspaces, while keeping the coherence within them intact.
We will consider only noise where the dephasing is in the canonical basis with a specific focus on two cases of this type of noise.

\subsection{Coupled noise}

In the first case, which we shall call \emph{coupled noise,} the $\hi_0=\textrm{span}\, A_0$ where $A_0\subseteq \{\ket{j}\}_j$ is some subset of canonical basis states. This is, for example, the case of a register consisting of qubits, when one of the qubits is affected by the dephasing---let $j$ be the qubit that undergoes decoherence, then $A_0$ is the set of all canonical states that have in their binary notation the same value, let us say $0$, on $j$th position. Considering the more general case of multiple dephasing qubits is beyond the scope of this paper and studied cases will not cover it; some results for globally (on all qubits) applied noise on qubit registers can be found in Refs.~\cite{Azuma02,ShMoBi03,CoDeBuDe16}.

Mathematically, applying Eq.~(\ref{eq:noise}), the coupled noise splits the density matrix for a system state into four blocks with the split corresponding to the two subspaces, and affecting only the non diagonal blocks,
\begin{equation}
\dep_p(\varrho)=\dep_p\left(\begin{bmatrix}
\hat\varrho_{00} & \hat\varrho_{01} \\
\hat\varrho_{10} & \hat\varrho_{11}
\end{bmatrix}\right)=
\begin{bmatrix}
\hat\varrho_{00} & (1-p)\hat\varrho_{01} \\
(1-p)\hat\varrho_{10} & \hat\varrho_{11}
\end{bmatrix}.
\end{equation}
Here the $\hat\varrho_{ij}$ are submatrices of the original state on respective subspaces $\hi_0$ and $\hi_1$. The effect of $\dep_p$ can be written for density matrix elements $\ket j\bra k$ also as
\begin{equation}
\label{eq:dephasing}
\dep_p(\ket j\bra k)=\begin{cases}
\ket j\bra k & \text{if $\ket j, \ket{k}\in\hi_0$}\\
& \text{or $\ket j, \ket{k}\in\hi_1$,}\\
(1-p)\ket j\bra k & \text{otherwise.}
\end{cases}
\end{equation}

\subsection{Decoupled noise}

In the second case, the dephasing within the set of elements from the canonical basis $A_0\subseteq \{\ket{j}\}_j$ will be decoupled (we shall denote $[A_0]=\{j:\ \ket{j}\in A_0\}$ and, similarly, $[A_1]$). Unlike in the previous case, for each canonical state of $A_0$ the dephasing shall act independently, while subspace determined by $A_1$ stays unaffected (i.e., has a coupled decoherence with $\hi_0$). This removes the symmetrical treatment of $\hi_0$ and $\hi_1$ from the previous case.
The decoherence on $\hi_0$ is separated here into the \emph{canonical decoherences} on each vector of $A_0$. In particular, for each $j\in [A_0]$ we define local dephasing $\dep_p^{(j)}$ as given by Eq.~(\ref{eq:noise}) with $\Pi_0=\ket{j}\bra{j}$. From now on we suppose the same noise rate $p$ for all $j$'s.

An important property is that the canonical dephasings with respect to different $j$'s commute,
\begin{equation}
\dep_p^{(j)}\circ \dep_p^{(k)} = \dep_p^{(k)}\circ \dep_p^{(j)}
\end{equation}
for $j,k\in [A_0]$. This can be confirmed by a simple calculation.
The overall (but localized to $\hi_0$) dephasing is defined as composition of all the canonical dephasings,
\begin{equation}
\dep_p=\underset{j\in [A_0]}{\bigcirc} \dep_p^{(j)}.
\end{equation}
Invoking Eq.~(\ref{eq:dephasing}), we can now describe action of decoupled noise $\dep_p$ for all density matrix elements $\ket j\bra k$.

If $j=k$ or $j,k\in A_1$, then
\begin{equation}
\dep_p(\ket j\bra k)=\ket{j}\bra{k}.
\end{equation}
If $j\in A_0, k\in A_1$ or $j\in A_1, k\in A_0$, then
\begin{equation}
\dep_p(\ket j\bra k)=(1-p)\ket{j}\bra{k}.
\end{equation}
And, finally, if $j,k\in A_0$ and $j\neq k$,
\begin{equation}
\dep_p(\ket j\bra k)=(1-p)^2\ket{j}\bra{k}.
\end{equation}

The density matrix is then affected in the following way:
\begin{equation}
\dep_p(\varrho)=\dep_p\left(\begin{bmatrix}
\hat\varrho_{00} & \hat\varrho_{01} \\
\hat\varrho_{10} & \hat\varrho_{11}
\end{bmatrix}\right)=
\begin{bmatrix}
\hat\varrho'_{00} & (1-p)\hat\varrho_{01} \\
(1-p)\hat\varrho_{10} & \hat\varrho_{11}
\end{bmatrix},
\end{equation}
where 
\begin{equation}
\hat\varrho'_{00}=(1-p)^2\hat\varrho_{00}+p(2-p)\;\diag[\hat\varrho_{00}],
\end{equation}
which is $\varrho_{00}$ with unchanged diagonal elements and off-diagonal elements scaled by factor of $(1-p)^2$.

Practically, this type of noise can be present when performing, e.g., a quantum-walk search where some of the corresponding vertices might be ``damaged.'' In Ref.~\cite{HiBeFe03}, an interferometric interpretation of quantum walks is presented and this view can have a literal meaning in realization, where noises affecting laterally close states (vertices) might be present (see also Fig.~\ref{fig:star}).

\section{Grover Search with dephasing}
\label{sec:invariant}

In this section, we shall expand the model presented in Sec.~\ref{sec:GS} by introducing the noise into the pure unitary search. 
In our studied case, the oracle $f$ marks only a single element, which we refer to as \emph{the target.} Without loss of generality this element will be the very first one. The rest of the elements will be called \emph{normal}. Some of these elements will be affected by noise; these will be numbered from $2$ to $k+1$, so there will be $k$ such elements. The target element might, or might not be affected by the noise and we will consider both possibilities. The rest $M=N-k-1$ elements will be normal ones that are unaffected by the noise.

Furthermore, we consider the dephasing rate to be uniform (either it describes the case of larger affected subspace with coupled noise or, if it is decoupled, the rate is the same for all elements). We will, however, set different rates on the target and on noisy normal (nontargeted) states in our computations to be able to distinguish between different cases of target either being or not being affected by the dephasing.

As noted, the invariant subspace formalism (see Refs.~\cite{krovi,ReHiFeBu09}) cannot be used directly, as the dephasing not only takes the system out of the subspace, but even more, it destroys the purity of the state. In what we shall present, we identify an invariant subspace $\subspace$ within the density matrix formalism. It is similar in spirit to the identification of decoherence-free subspaces \cite{dfs,krovi2} but we are not trying to use the subspace to produce error-free evolution; our aim is to take it as given and try to understand the effect of the noise. To this end, the subspace will be defined as a span of a specific set of operators that will no longer have a clear physical interpretation. They will, however, still define a linear space invariant under both the unitary evolution $\uni$ and the dephasing $\dep$. As the usual initial state Eq.~(\ref{eq:groverinit}) is from the subspace $\subspace$, any subsequent state within the evolution will also lie in this subspace $\subspace$.

\begin{figure}
\begin{center}
\includegraphics[scale=.97]{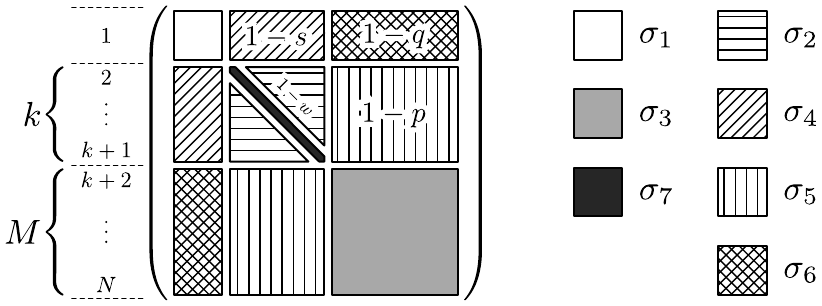}
\end{center}
\caption{The matrices forming the invariant subspace $\mathcal S$. Matrices $\sigma_1$, $\sigma_3$, and $\sigma_7$ have a non-zero trace, all others are traceless. The action of dephasing in the basis represented by these matrices is diagonal. The noise rates are written in the blocks for reference. Where none is present, a factor of $1$ is assumed.}
\label{fig:matrix}
\end{figure}

In the generality needed for the full analysis within the questions addressed by this paper, the subspace $\subspace$ will be seven-dimensional, $\subspace=\spann\{\sigma_j: j=1,2,\ldots,7\}$, where
\begin{align}
{\sigma_1} =& \ke 1\bra 1,\notag\\
{\sigma_2} =& \frac{1}{\sqrt{k(k-1)}} \sum_{j=2}^{k+1}\sum_{\substack{m=2\\m\neq j}}^{k+1}\ke j\bra m,\notag\\
{\sigma_3} =& \frac{1}{M} \sum_{j=k+2}^N\sum_{m=k+2}^N\ke j\bra m,\notag\\
{\sigma_4} =& \frac{1}{\sqrt{2k}} \sum_{j=2}^{k+1}(\ke 1\bra j+\ke j\bra 1),\notag\\
{\sigma_5} =& \frac{1}{\sqrt{2kM}} \sum_{j=2}^{k+1}\sum_{m=k+2}^N(\ke j\bra m+\ke m\bra j),\notag\\
{\sigma_6} =& \frac{1}{\sqrt{2M}} \sum_{j=k+2}^N(\ke 1\bra j+\ke j\bra 1),\notag\\
{\sigma_7} =& \frac{1}{\sqrt{k}} \sum_{j=2}^{k+1}\ke j\bra j.
\label{eq:sigmabasis}
\end{align}
We will call this also \emph{a $\sigma$-basis.}
The operators $\sigma_j$ are depicted in Fig.~\ref{fig:matrix} as different renormalized subparts of a uniform operator $\sum_{j,k}\ket{j}\bra{k}$.
This splitting recognizes the differences between the various parts of states (target/normal vertices and noise-affected/-unaffected) and it still allows us to define the unitary evolution $\uni$ on this subspace. Similarly, as in the Introduction, we use
\begin{equation}
t=\frac{2}{N},\qquad r=1-t=\frac{N-2}{N}.
\end{equation}

The unitary evolution $\uni$ is described by transformation rules within the subspace $\subspace$:
\begin{align}
\uni(\sigma_1)=& r^2\sigma_1+t^2\sqrt{k(k-1)}\sigma_2+t^2M\sigma_3-rt\sqrt{2k}\sigma_4\notag\\
         & +t^2\sqrt{2kM}\sigma_5-rt\sqrt{2M}\sigma_6+t^2\sqrt{k}\sigma_7,\notag\\
\uni(\sigma_2)=& t^2 \sqrt{k(k-1)}\sigma_1+[1+t(k-1)(tk-2)]\sigma_2\notag\\
         & +t^2M\sqrt{k(k-1)}\sigma_3+t(tk-1)\sqrt{2(k-1)}\sigma_4\notag\\
         & +t(tk-1)\sqrt{2M(k-1)}\sigma_5\notag\\
         & +t^2\sqrt{2Mk(k-1)}\sigma_6 +t(tk-2)\sqrt{k-1}\sigma_7,\notag\\
\uni(\sigma_3)=& t^2M(\sigma_1+\sqrt{k(k-1)}\sigma_2+\sqrt{2k}\sigma_4+\sqrt{k}\sigma_7)\notag\\
         &+(1-tM)^2\sigma_3 -(\sqrt{k}\sigma_5+\sigma_6)t(1-tM)\sqrt{2M},\notag\\
\uni(\sigma_4)=& tr\sqrt{2k}\sigma_1+\sqrt{2}t(1-tk)(\sqrt{k-1}\sigma_2+\sigma_7)\notag\\
         & -t^2M\sqrt{2k}\sigma_3-[r-tk(1-2t)]\sigma_4\notag\\
         & +t(1-2tk)\sqrt{M}\sigma_5+t(1-2t)\sqrt{Mk}\sigma_6,\notag\\
\uni(\sigma_5)=& t^2\sqrt{2Mk}\sigma_1+t(tk-1)\sqrt{2M}(\sqrt{k-1}\sigma_2+\sigma_7)\notag\\
         & +t(tM-1)\sqrt{2Mk}\sigma_3+t(2tk-1)\sqrt{M}\sigma_4\notag\\
         & +(2t^2Mk-r)\sigma_5+t(2tM-1)\sqrt{k}\sigma_6,\notag\\
\uni(\sigma_6)=& tr\sqrt{2M}\sigma_1-t^2\sqrt{2Mk}(\sqrt{k-1}\sigma_2+\sigma_7)\notag\\
         & -t(tM-1)\sqrt{2M}\sigma_3+t(1-2t)\sqrt{Mk}\sigma_4\notag\\
         & +t(1-2tM)\sqrt{k}\sigma_5 +(1-tk-2t^2M)\sigma_6,\notag\\
\uni(\sigma_7)=& t^2\sqrt{k}\sigma_1-t(2-tk)\sqrt{k-1}\sigma_2+t^2M\sqrt{k}\sigma_3\notag\\
        & +\sqrt{2}t(tk-1)(\sigma_4+\sqrt{M}\sigma_5)+t^2\sqrt{2Mk}\sigma_6\notag\\
        & +(1-2t+t^2k)\sigma_7.
\label{eq:7dimuni}
\end{align}
This means that the Grover evolution is restricted to the subspace $\subspace$ and the evolution in this subspace can be represented by a seven-dimensional unitary matrix $U$.

The noise, incorporated into the evolution as described in Eq.~(\ref{eq:generalevolution}), also keeps the subspace $\subspace$ invariant. Its representation within the subspace is given by a diagonal matrix described by the following transformation rules:
\begin{align}
\dep(\sigma_1)=&\sigma_1, & \dep(\sigma_2)=& (1-w)\sigma_2,\notag\\
\dep(\sigma_3)=&\sigma_3, & \dep(\sigma_4)=& (1-s)\sigma_4,\notag\\
\dep(\sigma_5)=& (1-p)\sigma_5, & \dep(\sigma_6)=& (1-q)\sigma_6,\notag\\
\dep(\sigma_7)=& \sigma_7. & &
\end{align}
The dephasing $\dep$ is described by four parameters $p$, $q$, $s$, and $w$ (we drop the index for the dephasing rate at this point in order to simplify the notation), which can be further restricted. In particular, when the noise is coupled, $w=0$. If the noise furthermore couples with the target element, $s=0$ and $q=p$. If the target is not coupled to the noisy normal vertices, $q=0$ and $s=p$. In the case of decoupled noise, we set $w=2p-p^2$ so that $1-w=(1-p)^2$. If, in this case the target is under the influence of the noise, $s=w$ and $q=p$. Otherwise we have $s=p$ and $q=0$. By proper choice of the parameters, we can thus simulate all relevant cases of how the dephasing affects the system.

Since both the unitary step $\uni$ and the dephasing $\dep$ keep the state from subspace $\subspace$ in the subspace, also their composition,
\begin{equation}
\label{eq:evolbasic}
\mathcal E(\varrho)=(\uni\circ \dep)(\varrho),
\end{equation}
which defines one step of the noisy evolution, will keep the state in $\subspace$.

It is worthwhile to note that the seven $\sigma_j$ matrices do not fully describe the whole Hilbert space. First, they have been chosen so that the symmetry within the sets of elements would be reflected. But, second, the span of these matrices, restricted to the state space, describes only real density matrices. This is enough for our purposes as both the unitary $\uni$ and the dephasing $\dep$ are described by real matrices, but it would not suffice for a general evolution.

To describe what happens to the initial state Eq.~(\ref{eq:groverinit}) under evolution Eq.~(\ref{eq:generalevolution}), we need to know how to represent the initial state in the $\sigma$-basis---every state from the subspace $\mathcal S$ can be expanded as a linear combination of the $\sigma_j$ matrices,
\begin{equation}
\label{eq:expansion}
\varrho_\va=\sum_ja_j\sigma_j\equiv \va\cdot\pmb\sigma
\end{equation}
with $\pmb\sigma$ being the vector of $\sigma_j$'s. Conversely, any state expressed as a combination of $\sigma_j$ matrices belongs to the invariant subspace $\mathcal S$ and, moreover, evolving it using the unitary $\uni$ or the dephasing $\dep$ will again produce a state from the invariant subspace $\mathcal S$. The initial state of equal superposition of all basis states  Eq.~(\ref{eq:groverinit}) is now given by vector
\begin{equation}
\label{eq:generalinit}
\va_{\mathrm{init}}=\frac{t}{2}\left(1,\sqrt{k(k-1)},M,\sqrt{2k},\sqrt{2Mk},\sqrt{2M},\sqrt{k}\right).
\end{equation}

In general, we can obtain coefficients $a_j$ from any state $\varrho_\va$ as presented in Eq.~(\ref{eq:expansion}) by defining inner product via the usual formula
\begin{equation}
\label{eq:innerproduct}
(\varrho_\va,\varrho_\vb)=\tr[\varrho_\va^\ast\varrho_\vb]= \sum_{j=1}^7a_j^\ast b_j.
\end{equation}
The coefficients are extracted by the formula
\begin{equation}
a_j=\frac{(\sigma_j,\varrho_a)}{(\sigma_j,\sigma_j)}=(\sigma_j,\varrho_a).
\end{equation}
The last equality holds due to orthonormality of $\sigma_j$'s under the inner product Eq.~(\ref{eq:innerproduct}).

The success probability we are interested in is given by the projection to the state $\sigma_1$ and so $p_{\mathrm{suc}}=a_1$.
Finally, the trace in the $\sigma$-basis is given as
\begin{equation}
\label{eq:trace}
\tr\varrho=a_1+a_3+\sqrt{k}a_7,
\end{equation}
since only $\sigma_1$, $\sigma_3$, and $\sigma_7$ are not traceless.

As the presented formulation in the seven-dimensional subspace $\subspace$ is still very demanding and to make an example of using presented identification of invariant subspaces, we look at the specific cases of the general evolution with the decoherence, starting from the simplest.

\section{Results}
\label{sec:results}

\subsection{Normal elements treated equally}

In this case, we will restrict ourselves to the case when all the normal elements of the database are the same---either not affected by the decoherence, or all under the influence of decoherence. Roughly speaking, we consider here the situation when $k\to N-1$ in which case the matrices $\sigma_3$, $\sigma_5$, and $\sigma_6$ are ill-defined and we do not include them in the computation any more. The unitary evolution is now described by transformation rules:
\begin{align}
\uni(\sigma_1)=& r^2\sigma_1+t\sqrt{2r(1+r)}\sigma_2\notag\\
& -r\sqrt{2t(1+r)}\sigma_4+t\sqrt{t(1+r)}\sigma_7,\notag\\
\uni(\sigma_2)=& t\sqrt{2r(1+r)}\sigma_1+(1-2rt)\sigma_2\notag\\
& +2r\sqrt{rt}\sigma_4-t\sqrt{2rt}\sigma_7,\notag\\
\uni(\sigma_4)=& r\sqrt{2t(1+r)}\sigma_1-2r\sqrt{rt}\sigma_2\notag\\
&+(2r^2-1)\sigma_4-\sqrt{2}rt\sigma_7,\notag\\
\uni(\sigma_7)=& t\sqrt{t(1+r)}\sigma_1-t\sqrt{t(1+r)}\sigma_2\notag\\
&+\sqrt{2}rt\sigma_4+r(1+t)\sigma_7.
\end{align}
The dephasing obeys
\begin{align}
\dep(\sigma_1)=& \sigma_1,\notag\\
\dep(\sigma_2)=& (1-p)^2\sigma_2,\notag\\
\dep(\sigma_4)=& (1-p)(1-q)\sigma_4,\notag\\
\dep(\sigma_7)=& \sigma_7.
\end{align}
We have therefore set $1-w=(1-p)^2$ and $1-s=(1-p)(1-q)$. This allows us to study three different scenarios:
\begin{enumerate}
\item[(A)] Broken target ($p=0,q\neq 0$)
\item[(B)] Global decoupled dephasing ($p=q$)
\item[(C)] Noisy normal vertices and unaffected target ($p\neq 0$, $q=0$)
\end{enumerate}
While both cases (A) and (C) treat the target in a different way than the normal vertices, case (C) seems to be rather unreasonable, as it would indicate, that while all elements undergo dephasing, the oracle does not. Case (A) is in this respect more reasonable, as it defines a system that evolves unitarily up to the oracle-selected element, which is noise-affected; this might, e.g., mean that the marking of the target is imperfect.

The initial state $\varrho_{\mathrm{init}}$ in the reduced basis is (here, for simplicity we identify state $\varrho$ with its vector $\mathbf a$ in the $\sigma$-basis)
\begin{equation}
\label{eq:initC2}
\varrho_{\mathrm{init}}=\frac{1}{2}\left(t,\sqrt{2r(1+r)},\sqrt{2t(1+r)},\sqrt{t(1+r)}\right).
\end{equation}

Unlike in the later cases (when the situation is treated analogously, but with more effort), we provide here a detailed analysis, starting with the properties of the unitary evolution and then treating dephasing as a small perturbation.

\subsubsection{Unitary evolution}
In the vector representation, the unitary evolution $\uni$ is given by a matrix whose columns are formed by the coefficients of corresponding $\sigma_j$ evolution. This matrix, $U$, has two double degenerate eigenvalues $1$ with eigenvectors
\begin{equation}
\nu_1=\frac{1}{\sqrt{2}}(\sqrt{t},0,0,\sqrt{1+r}),
\end{equation}
with overlap $1/\sqrt{N}$ with the initial state of equal superposition, and
\begin{equation}
\nu_2=\frac{1}{\sqrt{2(1+r)}}\left(\sqrt{r(1+r)},\sqrt{2},0,-\sqrt{rt}\right),
\end{equation}
with the overlap of $\sqrt{r/2}$ with the initial state of equal superposition. The two other eigenvalues are two complex-conjugated eigenvalues $\e^{\pm2\ii\theta}$, where $\cos\theta=r$, with eigenvectors
\begin{equation}
\nu_\pm=\frac{1}{2\sqrt{1+r}}(\sqrt{1+r},-\sqrt{2r},\pm\ii\sqrt{2(1+r)},-\sqrt{t})
\end{equation}
that have overlap of $\e^{\pm\ii\theta}/2$ with the initial state of equal superposition.

All the eigenvectors are normalized and mutually orthogonal under the definition of inner product Eq.~(\ref{eq:innerproduct}).
So, in the ideal case of no dephasing, the state after $m$ steps can be expressed as
\begin{equation}
\label{eq:eigenevolution}
\rho(m):=\uni^m\left[\sum_j (\nu_j,\varrho_{\mathrm{init}})\nu_j\right]=\sum_j (\nu_j,\varrho_{\mathrm{init}})\lambda_j^m\nu_j,
\end{equation}
where $\varrho_{\mathrm{init}}$ is the initial state Eq.~(\ref{eq:initC2}), $j$ indexes the eigenvectors $\nu_j$, and corresponding eigenvalues $\lambda_j$. Typically, the overlap $(\nu_j,\varrho_{\mathrm{init}})$ determines how much effect each eigenstate has on the evolution. Here the situation requires a more detailed analysis, as the probability of success is the quantity we consider.
The probability of success is given by the first element of used vectors and, therefore, the important eigenvectors are those having large value of $\omega_j:=|(\nu_j,\varrho_{\mathrm{init}})(\nu_j)_1|$.  In this specific example, eigenvector $\nu_1$ has overlap of $\omega_1\sim O(1/N)$, while all other eigenvectors have overlap $\omega_j\sim 1/2$. The evolution can be, after some manipulation, expressed by Eq.~(\ref{eq:GSsuccess}).

\subsubsection{Including the noise}

Now let us include also dephasing into our discussion.
With respect to the definition of trace in Eq.~(\ref{eq:trace}), only eigenvector $\nu_1$ has a nonzero trace. Since both the unitary $\uni$ and the dephasing $\dep$ are trace preserving, $\nu_1$ has to be the eigenvector of $\dep$ with the eigenvalue $1$ as well. This reduces the analysis to the three remaining eigenvalues---eigenvalue $1$ corresponding to the vector $\nu_2$ and eigenvalues $\e^{\pm2\ii\theta}$.

Already in this simple case, the analysis under full evolution is difficult and therefore we will restrict ourselves to the case of small values of $p$ and $q$ and use perturbation theory to find the corrections to the eigenvalues. As the overlaps of these eigenvectors with the initial state are of $O(1)$, under supposition of first-order terms of $p$ and $q$ these overlaps do not change significantly and we will treat them as constants as over the studied time of evolution they remain unchanged and do not change the success probability.

Note also that (even in the full problem) if we have no noise on normal elements ($p=0$), the 1-eigenvectors of $\uni$ are also eigenvectors of $\dep$, i.e., the corrections to the 1-eigenvalue terms will depend only on $p$---the noise rate on the normal elements, while the dependence on $q$ in the first-order approximation shall be absent.

The characteristic polynomial $P(\lambda)$ of $\mathcal E$ as defined in Eq.~(\ref{eq:evolbasic}) is of fourth order with one solution $1$. The other eigenvalue $1$ of the matrix $U$ is perturbed---setting $\lambda=1+\delta p$ we find that
\begin{equation}
\label{eq:ev1}
P(1+\delta p)=16p[N+(N-1)\delta]+R_2,
\end{equation}
where $R_2$ represents terms of higher order; these are of order $p^2/N$ and so are small for all considered $p$'s. Setting $\Delta P(1)$, being the first-order variation to $P$ at value $1$ we solve for $\Delta P(1)=0$ and get
\begin{equation}
\delta=-\frac{N}{N-1}, \qquad \tilde\lambda=1-\frac{N}{N-1}p,
\end{equation}
where we marked the approximated eigenvalue by tilde.

For the conjugated eigenvalues, we let $\tilde\lambda_\pm=\e^{\pm2\ii\theta}(1+\delta_\pm p+\gamma_\pm q)$, and solving for $\Delta P(\lambda_\pm)=0$ we get 
\begin{equation}
\tilde\lambda_\pm=\e^{\pm 2\ii\theta}\left[1-\frac{2N-3}{2(N-1)}p-\frac{q}{2}\right].
\end{equation}
The higher order terms are smaller than the leading terms if $p,q\ll 1/\sqrt{N}$. The full evolution is obtained from a formula similar to Eq.~(\ref{eq:eigenevolution}),
\begin{equation}
\rho(m)=\mathcal{E}^m\left[\sum_j (\nu_j,\varrho_{\mathrm{init}})\nu_j\right]\simeq\sum_j (\nu_j,\varrho_{\mathrm{init}})\tilde\lambda_j^m\nu_j,
\end{equation}
which to the first order of approximation gives
\begin{multline}
\label{eq:sucC2}
p_{\mathrm{suc}}(m)\simeq\frac{1}{N}+\frac{N-2}{2N}\left(1-\frac{N}{N-1}p\right)^m\\
-\frac{1}{2}\cos[(2m+1)\theta]\left[1-\frac{2N-3}{2(N-1)}p-\frac{q}{2}\right]^m\\
\simeq \frac{1}{2}\left[(1-p)^m-\cos[(2m+1)\theta]\left(1-p-\frac{q}{2}\right)^m\right].
\end{multline}
The second approximation is for $N\to\infty$.
We note that this result is consistent with the noiseless case of the Grover search, as for $p=q=0$ it gives Eq.~(\ref{eq:GSsuccess}).

\begin{figure}
\begin{center}
\includegraphics{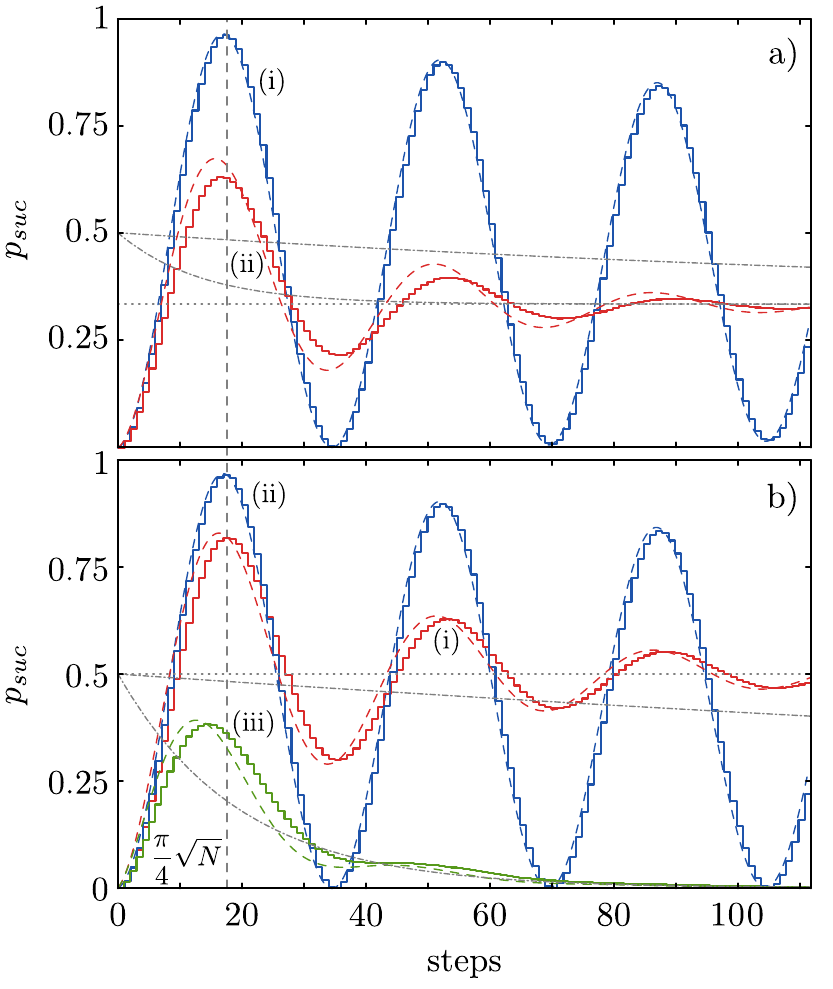}
\end{center}
\caption{Evolutions of the success probability (with $N=500$) under a) coupled noise---(a)(i) the periodic oscillations are pronounced when the target element is part of the (much) larger subset ($p=0.1, k=10, q=0$); the approximation works well. (a)(ii) The fit becomes worse and the probability grows more slowly in the limit of large times approaching $1/3$, when the size of the set where target belongs shrinks ($k$ increases to $N/2$). When the number of noisy elements in the normal set drops to $k=0$, we have only a noisy target, which is the same case as in the decoupled noise scenario (b)(i), with $p=0,q=0.05$. In this case, the limiting probability is $1/2$ and not $1/3$, as it is described by Eq.~(\ref{eq:gensuc}). A more general case is (b)(ii), where $k=10$ but without the noise affecting the target, which still offers a quadratic speedup. Again, the more elements are affected by noise, the easier it is to destroy the speedup. It gets even faster, when noise affects the target (b)(iii), where we show the limiting case of all elements under delocalized dephasing with $p=q=0.05$. The dashed grey line corresponds to the number of steps $m_0=\frac{\pi}{4}\sqrt{N}$, the dotted gray lines correspond to the linear terms of the corresponding $p_{\mathrm{suc}}$, and the dot-dashed grey lines to corresponding $p_{\mathrm{suc}}$ without periodic term.}
\label{fig:evolutions}
\end{figure}

Taking now $p=0$, $q\neq 0$, i.e., with just the target dephasing [case A, see Fig.~\ref{fig:evolutions}(b)(i)], the success probability has a stationary point $1/2$. The limiting state in this case is
\begin{equation}
\mu_0=\frac{1}{2\sqrt{1+r}}(\sqrt{1+r},\sqrt{2r},0,\sqrt{t}),
\end{equation}
which, with probability $1/2$ gives the target element and with probability $1/2$ projects into the normal subspace with equal probability to be located in any normal state. This case is studied also in Ref.~\cite{rastegin}, where exact results are presented. The reference contains some slight differences in the action of the noise, but it is possible to make a direct comparison with the results here. The noise used there is the same as in this paper but is used both after the oracle and the Grover unitary, and instead of reducing off-diagonal elements to $1-q$, it scales them by factor $\sqrt{\eta}$. As the noise commutes with the oracle, it can be rearranged to be used as in our case but twice in succession. Therefore we deduce identity $1-\eta=q$. With this substitution, the results of the reference fits our results.

After $m_0$ steps given by Eq.~(\ref{eq:m0}), when the cosine term becomes positive we have the success probability $p_{\mathrm{suc}}\geq 1/2$ which means, that the number of oracle calls by Eq.~(\ref{eq:avgstepsconst}) is still of order $\sqrt{N}$ and we do not lose the quadratic speedup.
However, this works only under given approximation ($q\ll 1/\sqrt{N}$)---outside this bound, the higher order terms will lead to the change of the frequency of oscillations and prolong the computation in a way that might eventually lead to the loss of quadratic speedup. This is indeed the case, as can be observed in Fig.~\ref{fig:scaling}.

On the other hand, additional dephasing $p\neq 0$ destroys the success probability quickly and the limiting state is the completely mixed state that gives probability of success only of order $1/N$. In this setting the only physically relevant situation is a global decoupled dephasing with $p=q$ [case B, see Fig.~\ref{fig:evolutions}(b)(iii)]. Invoking the results of Ref.~\cite{VrReReWo14}, we find out, that in this case the quadratic speed-up is lost as well and the number of oracle calls is of order $pN$, unless $p,q\ll 1/\sqrt{N}$. The same analysis holds also for the case C.

In this simplest example, the parameter regions for applicability were small and, in turn uninteresting, recovering only previous results. In the next section, we shall extend the model and show that in some particular cases the parameters regions can be extended beyond $1/\sqrt{N}$ bound and provide interesting results. As the computations get more involved, we remove lengthy expositions and present only results.

\subsection{Coupled noise on any subset of elements}
\label{sec:C3}

Splitting the set of normal elements into those affected by noise and those that are not increases the dimensionality of the problem. Considering only coupled noise does, on the other hand, allow the dimensionality to be only six, as the diagonal matrix $\sigma_7$ can be merged with the off-diagonal matrix $\sigma_2$ of affected normal vertices into one, 
\begin{equation}
\tilde\sigma_2=\frac{1}{k}\sum_{j=2}^{k+1}\sum_{m=2}^{k+1}\ket{j}\bra{m}.
\end{equation}
This state change affects several definitions from the introduction. First, the unitary transformation rules Eqs.~(\ref{eq:7dimuni}) will be different; under action of $\uni$ the matrix $\tilde\sigma_2$ evolves as
\begin{align}
\uni(\tilde\sigma_2)=& t^2k\sigma_1+(1-tk)^2\tilde\sigma_2+t^2kM\sigma_3-t\sqrt{2k}(1-tk)\sigma_4\notag\\
& -t\sqrt{2kM}(1-tk)\sigma_5+t^2k\sqrt{2M}\sigma_6.
\end{align}
All other transformation rules $\uni(\sigma_j)$ from Eqs.~(\ref{eq:7dimuni}) change only in the $\sigma_2$ terms. These changes can be collectively described by exchange rule
\begin{equation}
 \sigma_2 \to \sqrt{\frac{k}{k-1}}\tilde\sigma_2.
\end{equation}
The initial state is
\begin{equation}
\label{eq:initC3}
\varrho_{\mathrm{init}}=\frac{t}{2}\left(1,k,M,\sqrt{2k},\sqrt{2kM},\sqrt{2M}\right)
\end{equation}
and the trace Eq.~(\ref{eq:trace}) is now
\begin{equation}
\tr\varrho=a_1+\tilde a_2+a_3,
\end{equation}
where $\tilde a_2$ is the coefficient by $\tilde\sigma_2$.
The dephasing noise is acting according to the rules (displaying only those, that do not act as identity):
\begin{align}
\dep(\sigma_4)=& (1-s)\sigma_4,\notag\\
\dep(\sigma_5)=& (1-p)\sigma_5,\notag\\
\dep(\sigma_6)=& (1-q)\sigma_6.
\end{align}
This parametrization allows us to discern three particular physically interesting cases in this discussion (here we assume small $k$'s):
\begin{enumerate}
\item[(A)] In the coupled noise scenario with target lying in the larger subspace for which $q=0$ and $s=p$
\item[(B)] In the coupled noise scenario with target lying in the smaller subspace for which $s=0$ and $q=p$
\item[(C)] A specific case of decoupled noise when $k=1$ that can be obtained by setting $q=p$ and $s=2p-p^2$
\end{enumerate}
These cases will be analyzed below in separate subsections.
As the difficulty of solving the problem grows beyond the point when displaying of the intermediate calculations would be useful, we present only a simplified analysis.

The unperturbed problem now contains two pairs of complex-conjugate eigenvalues and a double-degenerate eigenvalue 1. One pair of complex-conjugate eigenvalues gives the usual periodic behavior. The second pair would introduce a different period to the evolution, but the two corresponding eigenvectors have zero overlap with our initial state and are, hence, unimportant in the analysis. In the eigenvalue-$1$ subspace, we can find a basis of two eigenvectors, one of which has trace zero. The other eigenvector is the only eigenvector with non zero trace and, thus, is also an eigenvector for the dephasing $\dep$. Success probability coming from this vector is $1/3$. The problem is now solved by perturbing the remaining eigenvector for the eigenvalue one and the relevant periodic part. 

\subsubsection{Coupled noise}

\begin{figure}[t]
\begin{center}
\includegraphics[scale=0.9]{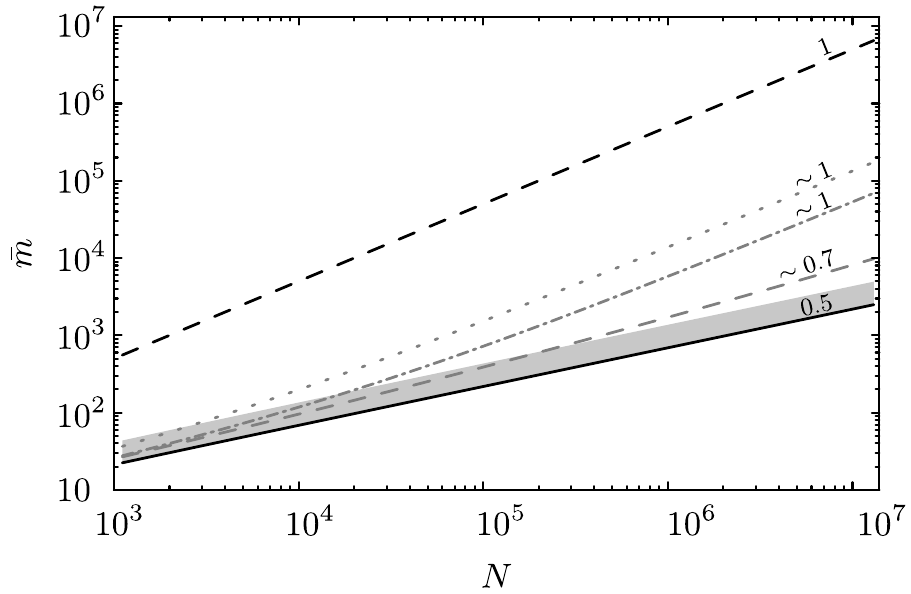}
\end{center}
\caption{Scaling properties of the average number of steps $\bar m$ given that a single experiment lasts $m_0$ number of steps (see explanation in the Appendix) for different situations in log-log scale as a function of the number of elemets $N$; these situations are displayed for coupled noise, however, qualitatively the same discussion holds also for decoupled noise. Black solid line depicts Grover search without noise, which has a quadratic speedup; dashed black line corresponds to the classical search. Cases having target noises $q\ll 1/\sqrt{N}$ and for normal vertices $kp\ll\sqrt{N}$, are retaining quadratic speedup (shaded region). Any deviation from these bounds leads to the loss of the speedup. Linear scaling is obtained whenever $q$ is constant (dot-dashed line); this loss is further pronounced when some noisy normal vertices are added (dotted line). Dashed gray line, when $k\simeq N^{0.7}$, diverges from quadratic speedup, but within the studied range it does not lead to linear scaling.
}
\label{fig:scaling}
\end{figure}

A situation (case A), when we have coupled noise and the target lying in the larger subspace, is obtained by taking $q=0$ and $s=p$. In this case the success probability is
\begin{multline}
\label{eq:approximation}
p_{\mathrm{suc}}(m)\simeq\frac{1}{3}+\frac{1}{6}\left[1-\frac{3kM}{(N-1)^2}p\right]^m\\
-\frac{1}{2}\cos[(2m+1)\theta]\left[1-\frac{k(2N-k-2)}{2(N-1)^2}p\right]^m.
\end{multline}
The approximations for the $1$-eigenvalue are appropriate for both $p\ll 1$ and $k\ll N$ and for the periodic term the applicability of the approximation is for $kp\ll\sqrt{N}$.
However, in this case, the former restriction on the $1$-eigenvalue term does not affect the complexity of the search, as with the constant term, this part of probability will always be $O(1)$; we are, hence, limitted only by constraint $kp\ll\sqrt{N}$. This is confirmed also numerically, see Fig.~\ref{fig:scaling}.

Within given constraints, after $m_0$ steps
given by Eq.~(\ref{eq:m0}) the periodic term is positive as well and the whole probability is of order one---the quadratic speedup is retained.
This situation is depicted in Fig.~\ref{fig:evolutions}(a)(i). However, if we, e.g., set $k=N/2$ [Fig.~\ref{fig:evolutions}(a)(ii)], which will take us outside the validity of the approximation, we are not guaranteed to have quadratic speedup any more.

Interestingly, we can look also at intermediate cases, when $k\sim N^\mu, \mu\geq 0.5$, where the numerical simulations up to $N$'s of size around one million show (see Fig.~\ref{fig:sizing}) that although the quadratic speedup is lost, the search can be still faster than the (linear) classical one, as the efficiency seems to scale as $N^\mu$. 

Note that taking $k\to 0$ recovers the Grover search Eq.~(\ref{eq:GSsuccess}) and taking $k\to N-1$ (which requires $p\ll1/\sqrt{N}$) recovers Eq.~(\ref{eq:sucC2}) in which the $p$ and $q$ have reversed roles, i.e., $p:=0$ and $q:=p$.
More importantly, comparing to the previous section we see the extension of validity of our approximations for small $k\leq\sqrt{N}$, for which we require only $p\ll 1$. That is, the approximation does not require the noise parameter to scale with the size of the system whenever $k$ is small.

Due to the symmetry of the coupled noise, the interpretation is that the quadratic speedup is retained for small noises whenever the target is in the larger noise-affected set. This symmetry also offers solution to case B, which is simply case A with $k\mapsto N-k-1$ and, hence, it does not require further analysis.

In both previous cases, the limiting state is the same, the single eigenvector with eigenvalue 1 of both $\uni$ and $\dep$,
\begin{equation}
\mu_0=\frac{1}{3}(\sigma_1+\tilde\sigma_2+\sigma_3),
\end{equation}
which has the same probability to be found in either of the three types of elements---target, noisy normal, and unaffected normal element.

\subsubsection{Decoupled dephasing on a target and single normal element}
\label{sec:C3special}

Case (C) is obtained by setting $q=p$ and $s=2p-p^2$ with $k=1$. Here we observe that the success probability is
\begin{multline}
\label{eq:C3special}
p_{\mathrm{suc}}(m)\simeq\frac{1}{3}+\frac{1}{6}\left[1-\frac{3(N-2)}{(N-1)^2}p\right]^m\\
-\frac{1}{2}\cos[(2m+1)\theta]\left[1-\frac{N^2-2}{2(N-1)^2}p\right]^m,
\end{multline}
where the periodic term is a good approximation only for $p\ll 1/\sqrt{N}$ valid for the periodic term. More interestingly, the middle term is no longer damped at rate $p$, but only at rate $p/N$, which does not have high importance now, but becomes valuable for general number of normal elements under decoupled dephasing.

\subsection{Decoupled noise on any subset of elements}
\label{sec:C4}

The most general case, as introduced in Sec.~\ref{sec:invariant} is also the most difficult to treat. The unitary evolution $\uni$ obeys the rules presented in Eqs.~(\ref{eq:7dimuni}) and the initial state is given by Eq.~(\ref{eq:generalinit}). As we have already treated coupled noise in the previous section, we reduce the number of noise parameters by setting the $\dep$ as follows:
\begin{align}
\dep(\sigma_2)=& (1-p)^2\sigma_2,\notag\\
\dep(\sigma_4)=& (1-p)(1-q)\sigma_4,\notag\\
\dep(\sigma_5)=& (1-p)\sigma_5,\notag\\
\dep(\sigma_6)=& (1-q)\sigma_6.
\label{eq:depC4}
\end{align}
All other rules act as identities.
This parametrization allows us to discern two cases---either target being affected by the noise ($p=q$ is assumed due to the symmetry considerations) or not affected ($q=0$, which is interesting for the case of small $k$'s in particular).

The solution of the unperturbed problem has, in comparison to the cases mentioned in Sec.~\ref{sec:C3}, a threefold degeneracy of eigenvalue 1, where we can find a third eigenvector for which the eigenvector has a zero overlap with the initial state and is not eigenvector for the dephasing $\dep$. Interestingly, all three eigenvectors are for $p=0$ eigenvectors of the dephasing $\dep$; the first-order dependence of the eigenvectors is thus independent of $q$.

With the analysis of the eigenvectors and known limit for $k=1$ (Sec.~\ref{sec:C3special}), we can obtain the first-order approximation to the success probability in the form
\begin{multline}
\label{eq:gensuc}
p_{\mathrm{suc}}(m)\simeq\frac{1}{k+2}+\frac{k}{2(k+2)}\left[1-\frac{(N-2)(k+2)}{(N-1)^2} p\right]^m\\
-\frac{1}{2}\cos[(2m+1)\theta]\left[1-\frac{k(2N-3)}{2(N-1)^2}p-\frac{q}{2}\right]^m.
\end{multline}
The strongest restriction on validity comes from the periodic term approximation, namely, $kp\ll\sqrt{N}$ and $q\ll1/\sqrt{N}$. This shows that whenever the target is affected by noise ($q>0$), the quadratic speedup is quickly lost as the threshold scaling $1/\sqrt{N}$ is very restrictive; this uncovers the destructiveness of the noise usually observed in the literature on the topic.

An interesting situation appears when $q=0$, i.e., when the target is in the noise-unaffected subspace. If the affected subspace is small ($k\leq\sqrt{N}$), the approximation to the success probability Eq.~(\ref{eq:gensuc}) is valid even for larger noise rates $p$. In other words, when the previous conditions are met, the search is robust toward the noise. The evolution in this case is depicted in Fig.~\ref{fig:evolutions}(b)(ii).

We can observe that the first static term decreases with growing $k$, but for small enough $k$'s still provides a quadratic speedup with scaling $k\sqrt{N}$ as after $m_0$ steps the periodic term becomes positive; see Appendix for explanation. For larger $k$'s approaching $\sqrt{N}$, the first term becomes small and does not allow quadratic speedup any more. On the other hand, the second term is damped only at a rate proportional to $kp/N$, which still provides a quadratic speedup by Eq.~(\ref{eq:avgstepssqrt}). For larger $k$'s, especially when approaching $N$,  we lose the speedup and the number of oracle calls grows towards $pN$; the solution approaches Eq.~(\ref{eq:sucC2}). In addition, Eq.~(\ref{eq:C3special}) can be recovered from Eq.~(\ref{eq:gensuc}) by taking $k=1$ and $p=q$.

Summing up the previous results, we can observe that the quadratic speedup of the quantum search can be retained when the noise affects a small number of normal elements, but not the target. In fact, our numerical simulations in Fig.~\ref{fig:scaling} show that the bound for the noise rate on the target, $q\ll 1/\sqrt{N}$, and the bound for the noise on normal vertices, $kp\ll\sqrt{N}$, is a rather strict one. Any deviation seems to lead to the loss of quadratic speedup if not any speedup. Probably the most disastrous is noise on the target, which very quickly leads to linear scaling. Having noise on the normal vertices then only intensifies the loss of the speedup.

\begin{figure}[t]
\begin{center}
\includegraphics[scale=0.9]{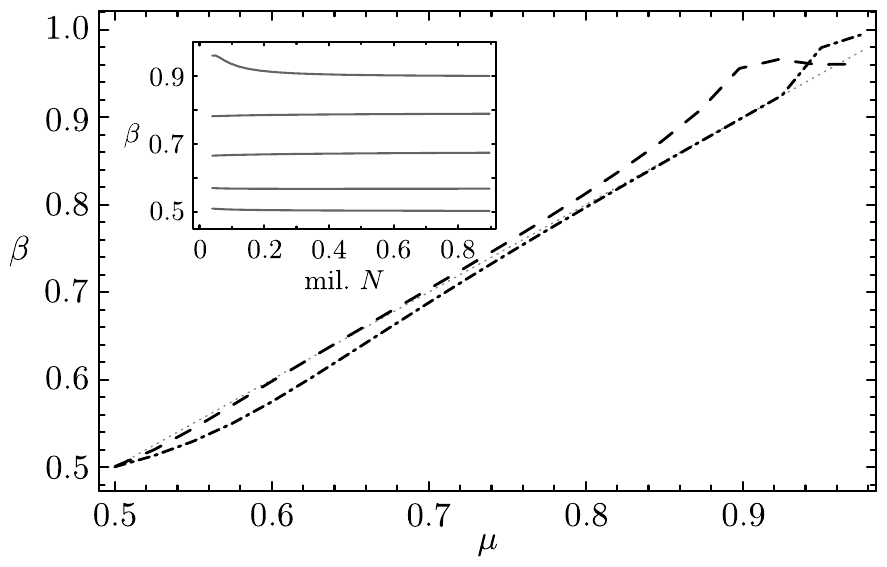}
\end{center}
\caption{
Dependence of efficiency exponent $\beta$, such that~$\bar m=O(N^\beta)$, as a function of exponent $\mu$ for the size of affected subspace, $k\simeq N^\mu$ for coupled dephasing (dot-dashed) and decoupled dephasing (dashed) is roughly linear. Inset shows the convergence of $\beta$ for various choices of $\mu=0.5,0.6,0.7,0.8,0.9$ from bottom to top.
}
\label{fig:sizing}
\end{figure}

If, however, the target is unaffected by the noise, the quadratic speedup is retained even for large noises $p$ and number of affected normal vertices up to $k\simeq\sqrt{N}$. Any larger scaling of $k\simeq N^\mu$ with $\mu>0.5$ seems to lead to the loss of the quadratic speedup. As seen in Fig.~\ref{fig:sizing}, the efficiency slowly deteriorates roughly as $N^\mu$. The convergence seems to be robust as seen in the inset of the figure.
Interestingly, these qualitative results hold irrespective of what type of dephasing we use, whether it is coupled, or decoupled noise.

In the following section, we shall make some parallels with quantum walks, where this method might be applicable as well.

\section{Applications to quantum walks}
\label{sec:qws}

The possibility of applying quantum searches within quantum walks does not provide additional results but the different point of view might be useful.
Quantum walks recently gained a lot of attention, as they are relatively easy to experimentally realize and, hence, may pose as a good testing ground for quantum algorithms that are applicable to them. One of the more prominent applications of algorithms to quantum walks are searches that have been described on various forms of graphs \cite{ShKeWh03,AaAm05,ReHiFeBu09}. Yet again, their results suffer from the problem of being proven to work only for the ideal unitary evolutions. The role of noise, which is important in experimental realizations, is only sparsely studied beyond observation of general properties on lattices \cite{KeTr03,BrCaAm03,AlRu05}.
Especially in the mentioned results, the theory of invariant subspaces, presented in Refs.~\cite{krovi, krovi2} in great depth, plays a deep role and in fact, was the original inspiration for this paper. For these reasons, the paper should not only offer results for the general quantum searches, but also more specifically for the community of quantum walks.

\begin{figure}
\begin{center}
\includegraphics[scale=0.7]{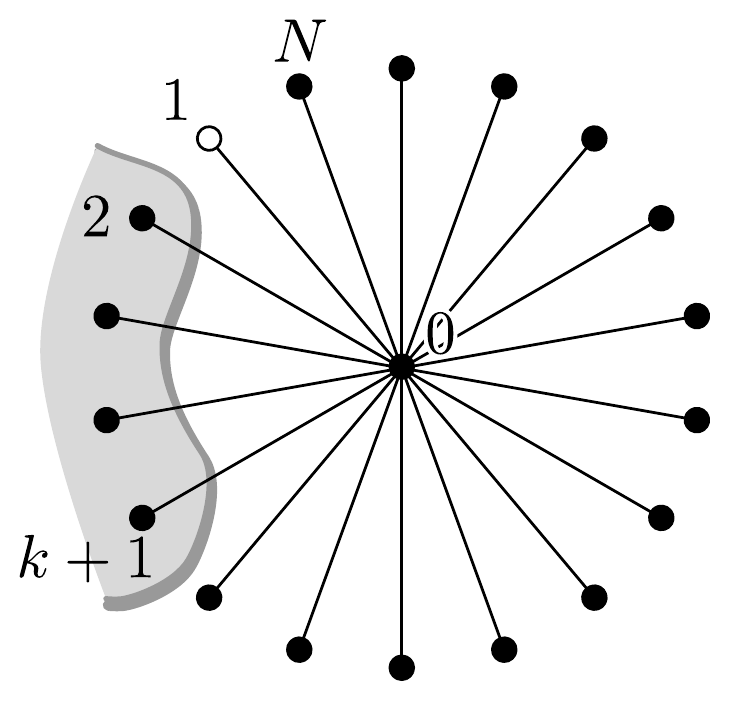}
\end{center}
\caption{\label{fig:star}Star graph with central vertex $0$, spokes' end-vertices $1$ through $N$. Vertex $1$ is marked by the oracle and vertices $2$ through $k+1$ are undergoing decoherence. When $k$ is small, the search by quantum walks can still be quadratically faster than in the classical case.}
\end{figure}

Quantum walks are quantum evolutions defined on graphs of specific structure. Following the interferometric interpretation of dicrete-time quantum walks introduced in Ref.~\cite{HiBeFe03}, the quantum walk is defined on a graph $\mathcal G=(V,E)$, where $V=\{1,2,\ldots,N\}$ is the set of vertices and $E$ is the set of edges, i.e., pairs of vertices. The graph is considered to be undirected, meaning, that if $(x,y)\in E$, then also $(y,x)\in E$. Quantum walk is then defined on a Hilbert space $\hi=\spann\{\ket{x,y}:(x,y)\in E\}$. The walker in these types of quantum walks can thus be interpreted as a particle traveling on edge $(x,y)$ from vertex $x$ to vertex $y$---we shall call these states \emph{edge states.}

The evolution $U$ on such a graph is defined via local evolutions on all vertices $x$, $U_x: \Omega_x\to A_x$, where
\begin{equation}
\Omega_x := \spann \{\ket{y,x}: y\in V, (y,x)\in E\}
\end{equation}
is the subspace of $\hi$ where the walker is traveling toward vertex $x$ and
\begin{equation}
A_x := \spann \{\ket{x,y}: y\in V, (x,y)\in E\}
\end{equation}
is the subspace of $\hi$ where the walker is traveling away from vertex $x$. The overall evolution $U$ is then a direct sum of all these sub-evolutions,
$U=\oplus_{x\in V}U_x$.

Quantum walks provide a playground for many quantum algorithms and quantum searches in particular. When performing a search with quantum walks, a usual assumption is homogenity of a graph (high symmetry) and also high homogenity of the quantum evolution. In particular, in connection to the oracle $f$ it is usually assumed that $U_x=(-1)^{f(x)}U_0$ where $U_0$ is chosen to be the inversion about average operator $G$, or the identity $I$ (depending on the setup). In these cases, the theory of Ref.~\cite{krovi} can be used to show quadratic speedups in localization of elements (vertices) for which $f$ gives one. 

If we want to include noise, considering localized decoherence has a straightforward interpretation as a spatially localized region where decoherence acts. Having, for example, a set of vertices $\mathcal M$ lying in this region, we can imagine noise acting on all edge states that originate in this region, e.g.
\[
A_0=\{|x,y\rangle:\ x\in\mathcal M,\ (x,y)\in E\}.
\]
In the same way as in the Grover search case studied before, we have a high symmetry imposed by the graph structure and, additionally, a splitting into noise-affected and noiseless parts. As in the previous case, we can look for invariant subspace for both the unitary evolution and the noise. 

As an example, let us consider a quantum-walk based search on a star graph (see Fig.~\ref{fig:star}) having one central vertex (labelled $0$) and $N$ vertices at the ends of the spokes (numbered from $1$ to $N$) that are connected to the central vertex. Let the oracle $f$ select one of the outer vertices, which we can set to be $1$ without loss of generality. In an ideal case, the evolution on this graph is described by three different local unitaries. The central vertex acts as the inverse about average Eq.~(\ref{eq:G}), i.e., $U_0=G$. The unmarked (normal) outer vertices simply reflect the walker back, i.e., $U_j=\id$ for $j=2,3,\ldots,N$. The target vertex $1$ reflects the particle back as well, but with an additional phase, $U_1=-\id$. Let us again use the notation $\uni(\varrho)=U\varrho U^\dagger$, and let the initial state be the equal superposition of outgoing walkers,
\begin{equation}
\label{eq:initQW}
\ket{\psi_{\mathrm{init}}}=\frac{1}{\sqrt{N}}\sum_{j=1}^N\ket{0,j}.
\end{equation}
What we shall show is that this problem can be mapped to our previous results for noisy quantum search.

Defined unitary evolution is an ideal situation, which in real experiment might be disturbed by noise that will again be labeled as $\dep$. Suppose some subset of vertices, from $2$ through $k+1$, is faulty and acts similarly as in Ref.~\cite{KoBuHi06}. 
In particular (e.g., due to thermal fluctuations), a random phase-shift $\ke{0,x}\mapsto\e^{\ii\phi}\ke{0,x}$ is introduced to each of the faulty vertices, where $\phi$ is sampled from a probability density function $\pi$ that is symmetric around $0$ and on interval $[-a,a]$. This phase-shift is considered to be dynamic, i.e., randomly sampled at each step.
This situation is also depicted in Fig.~\ref{fig:star}.

We can simplify the problem by considering two steps of evolution. Due to the nature of the initial state Eq.~(\ref{eq:initQW}) and the noise not affecting states of form $\ket{j,0}$, the application of the noise after every use of the unitary $\uni$, the noise $\dep$ will have no effect half of the time. In particular, when the state $\varrho$ is described as a particle leaving vertex $0$, then we have identity
\begin{equation}
(\uni\circ\dep)^2(\varrho)=(\uni^2\circ\dep)(\varrho).
\end{equation}

Furthermore, as we are unaware what the actual phase-shift is, to our best knowledge the state changes under the channel that is the average over all phase shifts:
\begin{equation}
\tilde\dep(\varrho)=\int_{-a}^a \pi(\phi) \dep(\varrho) d\phi.
\end{equation}
So, in the end, what we are interested in is evolution described at every step by $\mathcal E=\uni^2\circ\tilde\dep$. Let us now have a closer look at the introduced noise $\tilde\dep$ in this averaged image.

If we describe the quantum walker state by the density matrix, then we have three types of coefficients $\varrho_{j,m}$. First are those where both $j,m$ are not from the noise-affected subspace, or when $j=m$. In this case, the corresponding density matrix coefficients do not change under decoherence. If one of $j,m$ belongs to the affected subspace, but the other does not, the coefficent after the application of the noise $\dep$ acquires a phase $-\phi_j$ or $\phi_m$. In both cases, the averaging gives a multiplication factor to the coefficient
\begin{equation}
\label{eq:dephfromphase}
0\leq 1-p\equiv\int_{-a}^a \pi(\phi) \e^{\ii\phi} d\phi = 2\int_0^a \pi(\phi)\cos(\phi) d\phi\leq 1.
\end{equation}
Finally, if both vertices $j,m$ are from the affected subspace and $j\neq m$, then the density matrix coefficient acquires a phase $\phi_j-\phi_m$. Similarly as before, the averaging gives a prefactor to the coefficient of the form $(1-p)^2$.

We can observe now that this behavior of the decoherence $\tilde\dep$ is the same as the dephasing noise used in Eq.~(\ref{eq:depC4}). Furthermore, the unitary part of the evolution given by $\uni^2$ is described by transformation rules Eqs.~(\ref{eq:7dimuni}) where in the definition of the $\sigma$ basis Eqs.~(\ref{eq:sigmabasis}) we identify the edge states $\ket{0,j}$ with states $\ket{j}$. Hence, we have mapped the quantum-walk search problem with dynamical phase-shifting on some elements to the case of noisy quantum search from Sec.~\ref{sec:C4}. We can now apply the results obtained there also to this case.


\begin{table*}[t]
\setlength{\tabcolsep}{12pt}
\begin{tabular}{llll}
\hline
Noise &  Case &  Efficiency & Conditions\\ \hline\hline
&  target in the larger subspace & $O(N^{1/2})$ & $kp\ll\sqrt{N}$  \\
Coupled & target in the smaller subspace & $O(N^{1/2})$ & $p\ll1/\sqrt{N}$\\
& target in the larger subspace with $k\sim N^\mu$ & $N^\mu$ & $\mu\in[1/2;1]$ (numerical)\\
\hline
& noiseless target & $O(N^{1/2})$ & $kp\ll\sqrt{N}$  \\
Decoupled & noisy target & $O(N^{1/2})$ & $p\ll1/\sqrt{N}$\\
& noiseless target with $k\sim N^\mu$ & $N^\mu$ & $\mu\in[1/2;1]$ (numerical)\\
\hline
\end{tabular}
\caption{Observed speedups for regions of validity in different studied cases. Numerically observed cases do not have validity regions and their efficiency is conjectured from observation.}
\label{theonlytable}
\end{table*}

\section{Conclusion}
\label{sec:conclusion}

We have studied a Grover search with one target element under a local partially dephasing channel using a method of invariant subspaces in the density matrix formalism. The analysis shows interesting behavior in several cases that still allow for a quadratic speedup. The dephasing we considered was of two sorts---coupled, where the whole subspace was affected collectively, and decoupled, where the canonical elements of affected subspace were affected individually. The results are summarized in Table \ref{theonlytable}.

In the case of decoupled noise the Hilbert space is split into two parts, and the noise dephases these two parts. This means that the two subspaces are in a symmetrical position. If the target element is  in the smaller group, the dephasing has a very detrimental effect and, to retain a quadratic speedup, its rate $p$ needs to obey $p\ll1/\sqrt{N}$. If, however, the target element lies in the larger subspace and the size of the smaller subspace is $k$, then the condition for retaining a quadratic speedup is $kp\ll\sqrt{N}$.

In the case of decoupled noise, the symmetry between the spaces is broken, as the elements in the affected subspace are affected individually. If the target is part of this subspace, the noise destroys speedup very quickly and one can have a quadratic speedup only if $p\ll1/\sqrt{N}$. If, however, the target element is not in the noise-affected subspace, then to retain a quadratic speedup, it is again sufficient to fulfill condition $kp\ll\sqrt{N}$.

We have thus found regions of noise where the conditions are favorable for retaining the quadratic speedup---the condition is that the size of the subspace affected by noise should be smaller than $~\sqrt{N}$. Numerically, we have also observed that if the size of this region grows as $N^\mu$ with $\mu\in[1/2;1]$, the efficiency of the search tends toward the classical limit as $N^\mu$, which is worse than the best possible quadratic speedup, but still better than the classical bound. This has a consequence for a physically relevant case of a qubit register with one of the qubits decohering ($k=N/2$). In such a case, our results suggest that any speedup might already be lost unless the noise scales like $1/\sqrt{N}$.

\section*{Acknowledgements}
D.R. was financed by SASPRO Program No.~0055/01/01, co-funded by the European Union and Slovak Academy of Sciences. D.R. also acknowledges support from VEGA 2/0173/17 Project MAXAP, APVV-14-0878 Project QETWORK, Czech Grant Agency (GA\v CR) Project no.~GA16-22211S, and the support from Fulbright Commision.

\appendix

\section{Measuring the speed of the search}

When trying to decide how fast a search is (measured in the number of oracle calls), in general the usual approach to wait until the probability hits one may not work as quite commonly the success probability will be dampened fast enough not to get close to one. After an unsuccessful search, one needs to try again, which prolongs the search. In such case, one weighs different factors in---the smaller number of oracle calls one makes can lead to smaller success probability (at least right after the start of the search) but, on the other hand, the longer we let the system evolve, the higher contribution to the overall number of oracle calls it has in the end.

A good measure of how an algorithm can search for an element is then the compromise between the two drawbacks---the smallest average (expected) time to find a searched item. If the success probability of finding the item in $m$ steps is $p(m)$, then the average number of steps one needs to perform is
\begin{equation}
\label{eq:barn}
\bar m(m)=p(m)(m+1)\sum_{r=1}^\infty [1-p(m)]^{r-1}r=\frac{m+1}{p(m)},
\end{equation}
where $r$ counts the repetitions and we also account for an extra oracle call at the end, checking whether we have the correct item. In proving this, we used
\begin{equation}
\frac{1}{(1-q)^2}=\frac{d}{dq}\left[\sum_{r=0}^\infty q^r\right]=\sum_{r=1}^\infty rq^{r-1}.
\end{equation}
The optimal number of steps is found in the global minimum of function $\bar m$, which in particular is easy to find in numerical simulations. To find the optimal number of steps analytically, we could look at the maximum of $\bar m(m)$ when $\bar m'(m)=0$; this would give us condition
\begin{equation}
m+1=\frac{p(m)}{p'(m)}.
\end{equation}
This is usually hard to compute as it often leads to transcendental equations. However, in our cases when the success probability is given generally as
\begin{equation}
p_{\mathrm{suc}}(m)=\alpha+\beta(1-b)^m-\gamma(1-c)^m\cos[(2m+1)\theta],
\end{equation}
$\alpha,\beta,\gamma\geq0$, we can always choose $m:=m_0/2\equiv \pi\sqrt{N}/8$, which is a point when $(2m+1)\theta$ becomes closest to $\pi/2$ and the cosine contribution becomes positive. 
Thus, our success probability is
\begin{equation}
p_{\mathrm{suc}}(m)\geq\alpha+\beta(1-b)^{m}.
\end{equation}
When $\alpha$ is $O(1)$, we can immediately use Eq.~(\ref{eq:barn}) to find
\begin{equation}
\label{eq:avgstepsconst}
\bar m(m)\leq\frac{m+1}{\alpha}=O(\sqrt{N}).
\end{equation}

Once $\alpha$ becomes too small, possibly scaling with $N$, we shift our interest to
\begin{equation}
p_{\mathrm{suc}}(m)\geq\beta(1-b)^{m},
\end{equation}
while we suppose that $b=p/N^\kappa$ with $p\in [0,1]$ and $\kappa\geq 1/2$. Under these restrictions, the function
\begin{equation}
\left(1-\frac{p}{N^\kappa}\right)^{\sqrt{N}}
\end{equation}
is increasing in $N$ and so
\begin{equation}
p_{\mathrm{suc}}(m)\geq\beta\left(1-\frac{p}{4^\kappa}\right)^{\frac{\pi}{4}\sqrt{4}}\geq\frac{\beta}{3}.
\end{equation}
Using Eq.~(\ref{eq:barn}), we see that
\begin{equation}
\label{eq:avgstepssqrt}
\bar m(m)\leq\frac{3}{\beta}(m+1)=O(\sqrt{N}),
\end{equation}
where the last equality holds for $\beta=o(1)$.

On the other hand, when we take $b,p=O(1)$ (and when $\alpha$ is small), which is the usual form of restriction in our paper, we simply have
\begin{equation}
\bar m(m)\geq\frac{m+1}{2\beta(1-p)^{\sqrt{N}}}\geq\frac{m}{2\beta}(1+\sqrt{N}p)=O(pN).
\end{equation}
This means, that once $b$ is large and does not change with $N$, the search becomes inefficient and the number of oracle calls becomes linear in both $N$ and $p$ and the quadratic speedup is lost.
\end{document}